# Particle Refraction, Reflection and Channeling by Laser Beams

X. Artru[1], K.A. Ispirian[2]* and M.K. Ispiryan[3]

1) *Institut de Physique Nucleare de Lyon, CNRS/IN2P3 and Universite Claude-Bernard Lyon-1*
2) *Yerevan Physics Institute, Yerevan, Armenia (corresponding author; email: karo@mail.yerphi.am)*
3) *University of Houston, Houston, TX, USA*

**Abstract**

It is shown that the charged particles are refracted and reflected on the boundary of field free and laser field regions in vacuum. Simple and transparent estimates are given which show the possibility of channeling of charged and neutral particles having polarizability by strong electromagnetic field of certain laser bunches just as by the field of orientated crystalline planes and axes. These processes can be applied for production of femtosecond sliced electron bunches, for measurement their length and particle distribution as well as for production of femtosecond X-ray and terahertz pulses using transition, channeling and other types of radiation.

1. **Introduction**

The possibility of handling of the charged high energy particles by channeling [1], by recently discovered reflection [2], etc, by the strong atomic fields of the order of ~$10^8$ V/cm of crystallographic planes is well known. However little is known on the application of the stronger electromagnetic fields of intense laser beams for the same purposes though there is a progress in the field of advanced methods of particle acceleration [3] and for some other purposes.

The reflection of low energy electrons from standing light waves has been predicted by Kapitza and Dirac [4] and has been observed using laser beams in the work [5]. The possibility of reflection of electrons by evanescent waves produced during total reflection of femtosecond light pulses from the interface between dielectric and vacuum has been considered in [6]. Considering the motion of electrons in harmonic electromagnetic field due to Lorentz force after averaging over the field wavelength a simple, very important formula has been obtained (see formula (12b) of [6]) for the effective refractive index $n_e(\omega)$ of electron beam

$$n_e(\omega) = (1-\Delta)^{1/2} = (1-\eta^2/\gamma\beta^2)^{1/2}, \qquad (1)$$

where $\eta = eE/m\omega c$ is the laser field parameter (in the case other than circular polarization it is better to use $\eta^2 = e^2 \langle E^2 \rangle /(m\omega c)^2$ with average over the period of the electric field $E$ ) and $\gamma = 1/\sqrt{1-\beta^2}$ is the relativistic factor of the electron with velocity $v = \beta c$. From (1) it follows that for not very strong laser field parameters $\eta < 1$, when $\Delta < 1$, $n_e(\omega) < 1$ independently of the particle energy as for X-ray photons. Only for very high laser intensities, when $\eta > 1$, and $\gamma < \eta^2/\beta^2$, or when $\Delta > 1$, $n_e(\omega)$ becomes imagine which means that the electrons can not penetrate the field. Such a behavior of the effective refraction index $n_e(\omega)$ for electron beam reminds the variation of the refraction index $n_{ph}(\omega)$ for photons in vacuum due to the QED nonlinear effects in strong electromagnetic fields [7,8].

The averaged over the period force influences also the neutral particles. The physics reason for the acceleration or deflection of charged particles by laser fields in vacuum lies in the fact that any electromagnetic field with electric component $E$ acts on charged or neutral particles with a dipole force

$$F \approx \alpha \, grad \langle E^2 \rangle_{av}, \qquad (2)$$

where $\alpha$ is the particle dipole polarizability, and $\langle E^2 \rangle_{av} = I/8\pi c$ ( $I$ is the laser intensity) is the square of the electric field amplitude averaged over the field oscillation period. For atoms the values of $\alpha$ and $F$ are relatively large, and the gradient force (2) has wide application (see [9,10]). For electrons they are small, $\alpha = (-e^2/m\omega^2)$, nevertheless, one can use the dipole force for focusing low energy electron beams [11], vacuum acceleration in Gaussian laser beams, etc. Though the approach of introducing refraction index for electrons is beautiful and analytically tractable, it is some approximation because the laser fields are not of constant magnitudes and have no sharp boundaries, and numerical simulations are necessary [6,11]. Results close to those of [6, 11] one can, in principle, obtain by the method [12] of nonadiabatic tunneling in ponderomotive barriers since in localized regions of laser fields the particles are scattered elastically.

Let us note that recently the deflection of fast charged particle in electromagnetic field or refraction by forces other than the dipole forces has been considered in the works [13-16]. It has been shown that the interaction of an electron beam with polarized plane electromagnetic wave of laser photons propagating in the same direction in a finite interaction region results in significant periodical transversal deflection of the electrons after a relatively long field free region. In the case of linear and circular polarizations this deflection is, respectively, linear and circular, perpendicular to the initial electron beam direction. Solving the equation of motion it has been shown that this deflection can be used for femtosecond scanning, chopping (slicing) and measurement of length of electron bunches and construction of femtosecond oscilloscopes. The reason of such a deflection lies in the fact that on a finite length $L_{coh} \approx \lambda \gamma^2$ of interaction the electrons "feel" almost the same field of the co propagating laser photon beam.

In [17] the possibility of transmission, reflection and capture of electron beams by realistic Gaussian laser fields in vacuum has been shown numerically solving the equations of motion without averaging as in [13-16]. It is discussed how the more correct simulation results differ from ones obtained in plane wave approximation and shown that very strong, but available laser fields result in energy gain, i.e. longitudinal acceleration, during these processes. The results obtained in [13-16] in the approximation of plane electromagnetic waves aimed to obtain tranversal deflection with the help of weaker laser fields need to be confirmed by numerical simulations before beginning any experimental effort. In this work using (1) for the first time it is shown the possibility of reflection and refraction of charged particles entering into a region with constant laser field from vacuum as well as particle channeling in the field of a cylindrical laser beams.

## 2. Reflection and Refraction of Particles by Laser Beams

In this section we shall show phenomenologically that one can obtain reflection and refraction of electron beams entering from a field free into field filled region in vacuum directly without any evanescent field and material. Moreover it can be shown the reality of total inner (more correctly external) reflection. Indeed, (see Fig. 1) according to Snell's law when in vacuum an electron beam incidences with incident angle $\theta_{inc}$ or complementary "grazing" angle $\phi_{gr} = \pi/2 - \theta_{inc}$ onto a plane boundary of a region filled with laser photons with constant density, a large part of the electron beam will be refracted under larger refraction angle $\theta_{refr} = \arcsin(\sin\theta_{inc}/n_e)$. Since the difference between the refraction indices of two "media" is small, a small part of the electron beam will be also reflected from the boundary under angle $\theta_{refl}$ ($|\theta_{refl}| = |\theta_{inc}|$). However, when $\theta_{refr} \to \pi/2$, a total internal, or, more correctly, external reflection of the electron beam will take place when the beam incidences under grazing angles smaller than a certain critical grazing angle $\phi_{crit}$ given by the expression

$$\phi_{crit} = \arccos(n_e) . \qquad (3)$$

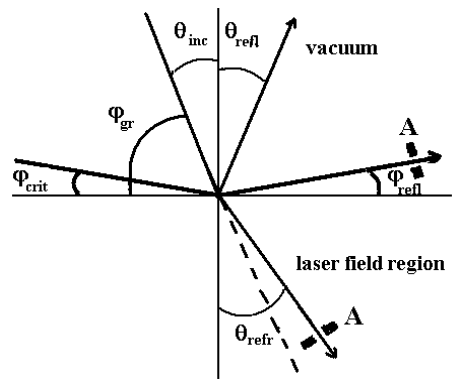

Fig.1 The electron beam refraction and reflection on the boundary laser-vacuum

For relativistic particles with $\gamma \gg 1$ and realistic laser intensities $\eta \leq 1$, when $\Delta \ll 1$, after Fourier expansion one obtains that the critical angle is proportional to $\eta$ and inverse proportional to $\sqrt{\gamma}$.

$$\phi_{crit} \approx \sqrt{\Delta} = \eta/\beta\sqrt{\gamma} . \qquad (4)$$

The dependence of $\phi_{crit}$ on kinetic energy of electrons for $\eta = 1$ is shown in Fig.2.

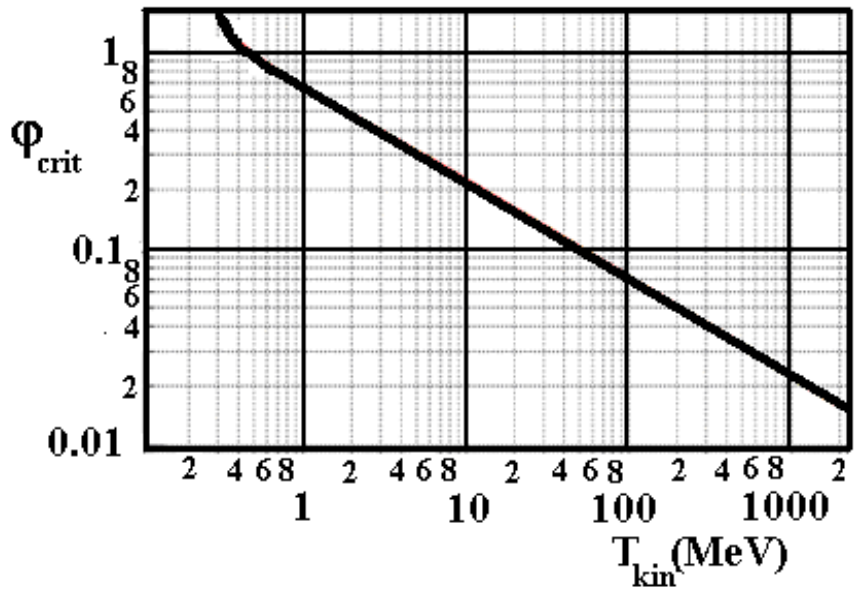

Fig.2. The dependence of the critical angle of total reflection $\phi_{crit}$ on electron's kinetic energy $T_{kin}$.

As it is seen the magnitude of the critical angle is significantly large for non relativistic energies and decreases with the increase of electron energy. Let us make some rough estimates. Let the laser beam be produced with the help of now available laser with $\lambda_L$ =800 nm, bunch length (rms) $\tau_L$ =50 fs or $L_L$ =15 $\mu m$, radius $R=50$, $\lambda_L=40 \mu m$, energy U=2.8 mJ and therefore $\eta=0.021$. For 20 MeV electron beam and such L-beam $\phi_{crit}=3.4$ mrad. However, since the real laser fields have no sharp boundary it is necessary to carry out numerical simulations as in [17] taking into account the real experimental conditions.

With the help of appropriate angular apertures (A in Fig 1) and femtosecond laser beams one can use such refraction or total reflection for slicing [18,19] of longer electron bunches into femtosecond electron bunches which further can be used for other purposes, for instance, for production of femtosecond X-ray bunches via the processes of transition, channeling, or other radiation mechanism. The answer of the question- what is more advantageous; refraction or reflection, TR or ChR for such application- depends on many available factors. One can measure the length of fs electron bunches by measuring the yield of the reflected or refracted bunches upon the delay between the e- and L-beams having the same frequency. Such fs measurements which remind the method using none linear optics of electrooptical (EO) crystals [20] can be carried out in different ways. Just as in the case of X-ray beams, one can use the refraction of electrons for focusing electron beams by making optical lenses.

### 3. Channeling of Charged Particles in Laser Beams

The channeling of particles in laser plasma channels has been studied as a method for particle acceleration [3]. Here we shall consider the channeling of particles in vacuum or laser field regions. Let an electron beam moves in the vacuum central part of a glass tube through the walls of which (see Fig. 3a) a short laser bunch is propagating in the same direction. Let us assume that there is a good synchronization between the e- and L-beams due to the optical refraction index of the glass tube. According to [6] if certain conditions between the parameters of e- and L-beams are fulfilled the electrons will be channeled as in the case of X-ray photons in capillaries [21] or charged particles in nanotubes [22] or photons in optical fibers with radial variation of the refraction index [23]. Now again without discussing the methods of production let us assume that we have tailored cylindrical laser beam with hollow (field free) axial part up to the radius $R_{in}$ and nearly constant photon density (see Fig. 3a) in a "laser tube" with inner and outer radii $R_{in}$ and $R_{out}$. According to the above said if the electrons of the beam enter into the hollow part of the "laser tube" under entrance angle $\theta_{entr} = \phi_{inc}$ with respect to the laser tube axis less than the above critical angle, then approaching the inner side of the sufficiently thick walls of the "laser tube" the electrons will be reflected and without energy loss will be channeled just as X-ray photons in capillaries [21]. The so called critical channeling angle is equal to the above given critical reflection angle (3) equal to $\theta_{entr} < 3.4$ mrad with respect to the laser axis.

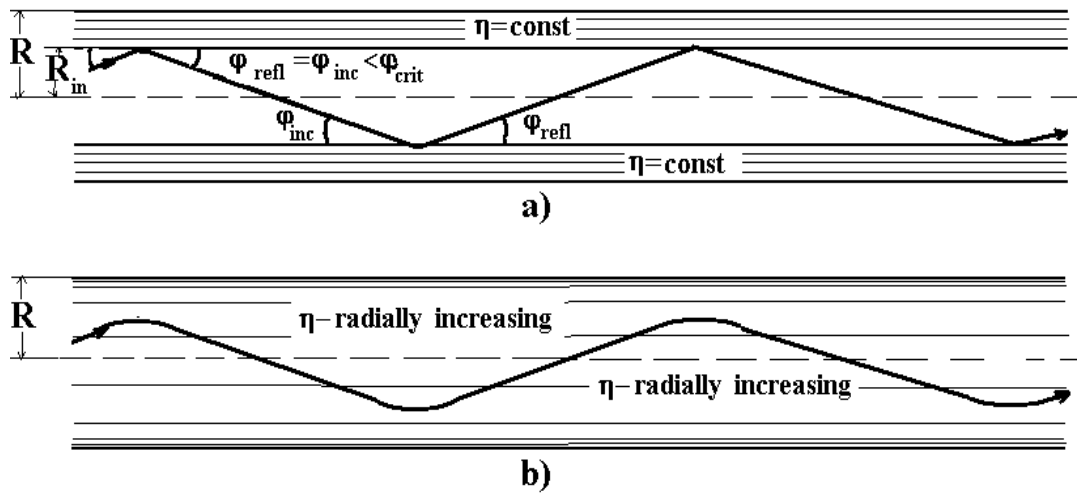

Fig. 3. The cross section of cylindrical laser beams a) hollow with constant photon distribution near the beam wall and b) with radially increasing photon density and the trajectories of channeled electrons with total reflections and smooth bendings, respectively.

Electron beam channeling will take place also, if in the transversal direction the laser beam photon density and the parameter, $\eta$, increases with the increase of distance $r$ from the axis (see Fig. 3b), i.e. in contrast to the usual Gaussian laser beams the photon density, therefore, also the laser intensity has an inverted Gaussian density distribution,. According to the above said an electron propagating under small angles with respect to the axis will experience an increasing expelling force directed to the laser beam axis when its $r$ increases. Instead of polyline trajectories in Fig.3a one has smooth trajectories. Therefore, when certain conditions take place the particle motion will be channeling motion. By the way, available Gaussian laser beams are required for channeling of positrons and this fact makes the experimental study easier. It is clear that the real trajectories will differ from those calculated in [13-16] in the approximation of plane waves. Let us remind that according to plane wave calculations [13-16] in the case of linear polarization of laser photons the trajectories are planar oscillations with very small amplitudes (less than the photon wavelength) and transversal drift depending on the initial phase of the electron entrance into the field region. In the case of circularly polarized laser beam the similar oscillatory trajectories of various electrons have gyration depending on the initial phase. The amplitude and period of these oscillations/girations are very small compared with the amplitude and period of motion of particles in magnetic undulators and approaches to corresponding amplitude and period in crystalline undulators. Again in the case of laser channeling to find the trajectories for the given parameters of e- and L-beams it is necessary to carry out numerical simulations similar to [17]. Similar processes take place during propagation of photons in optical fibers with radial variation of the refraction index [23], in the case of planar channeling between the crystallographic planes or better in the case of channeling of positive particles in carbon nanotubes [22] as well as in some sense in the case of laser beam self focusing [24] when due to QED nonlinear effects the refraction index for photons is varied in $r$ direction due to the variation of the density of laser beam.

## 4. Discussion

During the above description of the processes we have not taken into account many imperfectness, the energy losses and multiple scattering of electrons, the minimal thickness of the laser fields which still provides the above effects, etc. We have not discussed the methods of injection and ejection of particles into channels, of synchronization of e- and L-beams, which must be similar to the same methods applied for the particle acceleration. These and other problems as well as applications such as joint propagation of e- and L- beams in cosmos, joint laser and ion nuclear fusion, etc, will be considered in our further publications in which it will be given more accurate theoretical description of the above processes, similar to the Lindhard theory for particle channeling as well as numerical simulations which can finally prepare the experimental study of the processes.